\documentclass[aps,pra,superscriptaddress,reprint,floatfix,longbibliography,showpacs]{revtex4-1}
\usepackage[english]{babel}
\usepackage{amsmath}
\usepackage{amssymb}
\usepackage{graphicx}
\usepackage{hyperref}
\usepackage{upgreek}
\usepackage{epstopdf}
\usepackage{color}

\begin{document}
\title{High-pressure $versus$ isoelectronic doping effect on the honeycomb iridate Na$_2$IrO$_3$ }

\author{V. Hermann}
\affiliation{Experimentalphysik II, Augsburg University, 86159 Augsburg, Germany}
\author{J. Ebad-Allah}
\affiliation{Experimentalphysik II, Augsburg University, 86159 Augsburg, Germany}
\affiliation{Department of Physics, Tanta University, 31527 Tanta, Egypt}
\author{F.~Freund}
\author{I.~M.~Pietsch}
\author{A.~Jesche}
\author{A.~A.~Tsirlin}
\affiliation{Experimentalphysik VI, Center for Electronic Correlations and Magnetism, Augsburg University, 86159 Augsburg, Germany}
\author{J.~Deisenhofer}
\affiliation{Experimentalphysik V, Center for Electronic Correlations and Magnetism,
Augsburg University, 86135 Augsburg, Germany}
\author{M.~Hanfland}
\affiliation{European Synchrotron Radiation Facility, BP 220, 38043 Grenoble, France}
\author{P. Gegenwart}
\affiliation{Experimentalphysik VI, Center for Electronic Correlations and Magnetism, Augsburg University, 86159 Augsburg, Germany}
\author{C. A. Kuntscher}
\affiliation{Experimentalphysik II, Augsburg University, 86159 Augsburg, Germany}
\email{christine.kuntscher@physik.uni-augsburg.de}

\begin{abstract}
We study the effect of isoelectronic doping and external pressure in tuning the ground state of the honeycomb iridate Na$_2$IrO$_3$ by combining optical spectroscopy with synchrotron x-ray diffraction measurements on single crystals.
The obtained optical conductivity of Na$_2$IrO$_3$ is discussed in terms of a Mott insulating picture \textit{versus} the formation of quasimolecular orbitals and in terms of Kitaev-interactions.
With increasing Li content $x$, (Na$_{1-x}$Li$_x$)$_2$IrO$_3$ moves deeper into the Mott insulating regime and there are indications that up to a doping level of 24\% the compound comes closer to the Kitaev-limit.
The optical conductivity spectrum of single crystalline $\alpha$-Li$_2$IrO$_3$ does not follow the trends observed for the series up to $x=0.24$.
There are strong indications that $\alpha$-Li$_2$IrO$_3$ is less close to the Kitaev-limit compared to Na$_2$IrO$_3$ and closer to the quasimolecular orbital picture.
Except for the pressure-induced hardening of the phonon modes, the optical properties of Na$_2$IrO$_3$ seem to be robust against external pressure.
Possible explanations of the unexpected evolution of the optical conductivity with isolectronic doping and the drastic change between $x=0.24$ and $x=1$ are given by comparing the pressure-induced changes of lattice parameters and the optical conductivity with the corresponding changes induced by doping.

\end{abstract}

\maketitle

\section{Introduction}

The $5d$ transition metal compounds show strong spin-orbit coupling (SOC) concurrent with electronic correlations, leading to interesting electronic and magnetic properties. They are discussed in terms of various exotic ground states, like topological insulators~\cite{Shitade.2009, Pesin.2010, Yang.2010}, Mott insulators~\cite{Kim.2008, Kim.2009, Jackeli.2009, Watanabe.2010, Martins.2011}, Weyl semimetals~\cite{WitczakKrempa.2012, Go.2012, Sushkov.2015}, and spin liquids~\cite{Okamoto.2007, Singh.2013, Winter.2016, Alpichshev.2015}.
One important family among the $5d$ transition metal compounds are the 213 iridates $A_2$IrO$_3$ (with $A$=Na, Li), which have a honeycomb layered structure consisting of IrO$_6$ octahedra \cite{Choi.2012, Ye.2012, Winter.2017}.
Na$_2$IrO$_3$ was initially interpreted in terms of a topological insulator \cite{Shitade.2009}, however, experimental and theoretical investigations showed the Mott-insulating nature\cite{Sohn.2013, Comin.2012, Gretarsson.2013, Li.2015, Kim.2012} in the vicinity of a Kitaev spin liquid \cite{Winter.2016, Li.2017}. Also a band insulator picture with a quasimolecular orbital (QMO) ground state was suggested as an explanation for its insulating behavior \cite{Kim.2016, Foyevtsova.2013}.
Both Na$_2$IrO$_3$ and $\alpha$-Li$_2$IrO$_3$ order antiferromagnetically below $T_\text{N}\approx 15\,$K excluding the realization of a ``pure'' Kitaev spin liquid model \cite{Singh.2010, Singh.2012, OMalley.2008, Williams.2016}, while $T_N$ systematically decreases by substituting Na atoms by Li \cite{Manni.2014}.

The electronic properties of the iridates are dominated by the IrO$_6$ octahedra, where
the centered Ir$^{4+}$ ions with five $5d$ electrons are surrounded by six O$^{2-}$ ions. In Na$_2$IrO$_3$,
the octahedral crystal field (with a small trigonal distortion \cite{Kim.2014}) will largely split the Ir t$_{2g}$ and e$_g$ manifolds, so that all five
electrons occupy the t$_{2g}$ manifold. This t$_{2g}$ manifold is reconstructed into lower-lying, filled
$j_\text{eff}=3/2$ states and half-filled $j_\text{eff}=1/2$ states by the strong SOC of the heavy Ir.
The Coulomb repulsion $U$ splits the half-filled $j_\text{eff}=1/2$ band into an occupied lower
Hubbard band and an unoccupied upper Hubbard band, like in Sr$_2$IrO$_4$, leading to the
opening of a gap~\cite{Comin.2012, Sohn.2013, Gretarsson.2013} with the size $\approx 340\,$meV, being independent of temperature \cite{Comin.2012}.

The application of external pressure is a very efficient and clean way to tune the ground state of materials without introducing additional scattering centers, in contrast to chemical pressure which works via atomic substitution.
Pressure-dependent resistivity measurements on the perovskite 214 iridate Ba$_2$IrO$_4$ revealed an insulator-to-metal transition at a critical pressure of $13.4\,$GPa \cite{Okabe.2011}, while the related compound Sr$_2$IrO$_4$ showed a persistent non-metallic behavior up to a pressure of $55\,$GPa \cite{Zocco.2014}.
In contrast, most of the pyrochlore iridates $R_2$Ir$_2$O$_7$ (with $R$ being a rare earth element) are already metallic at ambient conditions with an insulating or semimetallic  phase at low temperatures \cite{Ueda.2012, Lee.2013, Sushkov.2015}.
Recently, indications for a pressure-induced quantum spin liquid state are observed for $\alpha$-RuCl$_3$ \cite{Wang.2017, Cui.2017}, which also orders in a honeycomb lattice similar to $A_2$IrO$_3$.

Here, we study the effect of external pressure and isoelectronic doping on the electrodynamical and structural properties of the honeycomb iridate Na$_2$IrO$_3$.
We further juxtapose these effects and discuss which of them is more promising for tuning the system towards the Kitaev limit.

\section{Methods}

(Na$_{1-x}$Li$_x$)$_2$IrO$_3$ single crystals (for $x\leq0.24$) were prepared by a solid-state synthesis as described previously \cite{Singh.2010, Manni.2014}.
Actual doping levels were determined by Laser Ablation – Inductively Coupled Plasma – Mass Spectrometry (LA-ICP-MS).
$\alpha$-Li$_2$IrO$_3$ single crystals were grown by vapor transport of separated educts as described in Ref. \onlinecite{Freund.2016} using elemental lithium and iridium as starting materials.
The samples were characterized by X-ray diffraction, specific heat, and magnetic susceptibility measurements in order to ensure phase-purity and crystal quality. No foreign phases were detected.

Room-temperature near-normal incidence reflectance spectra were measured on single crystals ($ab$ plane) in the frequency range $200-38000\,\text{cm}^{-1}$ ($0.025-4.7\,$eV) for Na$_2$IrO$_3$ and $\alpha$-Li$_2$IrO$_3$, and in the range $200-25000\,\text{cm}^{-1}$ ($0.025-3.1\,$eV) for Li-doped Na$_2$IrO$_3$. The Na$_2$IrO$_3$ and $\alpha$-Li$_2$IrO$_3$ single crystals were partially transparent, thus additional transmission measurements were performed.
The reflectance and transmittance measurements were carried out with a Bruker Vertex v80 Fourier transform infrared spectrometer in combination with an infrared microscope (Bruker Hyperion) with a 15$\times$ Cassegrain objective.
The intensity reflected from an Al mirror served as reference to obtain the absolute reflectance spectra.
For the transmittance spectra we measured the intensity $I_\text{s,trans}(\omega)$ of the radiation transmitted through the sample.
As reference the intensity $I_\text{ref,trans}(\omega)$ transmitted through air was used.

In case of the pure compounds Na$_2$IrO$_3$ and $\alpha$-Li$_2$IrO$_3$, a Kramers-Kronig analysis of the reflectance was combined with a direct analysis of the reflectance and transmittance ($R$+$T$ analysis), following Ref.\ \onlinecite{Xi.2013}, to obtain the real part of the optical conductivity, $\sigma_1$.
In the partial transparent range, the $\sigma_1$ of $R$+$T$ analysis is used, while in the opaque range the results from a Kramers-Kronig analysis are used
\footnote{From the $R$+$T$ analysis one can access the low-energy phonon modes for the partial transparent sample, which are too weak to be obtained by the reflectance measurements.}.
Since the measured single crystals of the doped compounds (Na$_{1-x}$Li$_x$)$_2$IrO$_3$ were opaque in the whole measured frequency range, the optical conductivity was obtained {\it via} Kramers-Kronig analysis of the reflectance data.
For the high-energy extrapolation \cite{[{Reflectance and reflectance phase were calculated in the range $10\,$eV$-30\,$keV from scattering functions $f_1$ and $f_2$ of }] [{with the use of a program written by Charles Porter.}] Henke.1993} we interpolated the frequency range between the measured data and the calculated spectra above $80000\,\text{cm}^{-1}$ ($10\,$eV) according to a power series in 1/$\omega^s$ with $s$ up to 4 \cite{Tanner.2015}.
The obtained $\sigma_1$ spectra were fitted with a simple Lorentz-oscillator model.

Pressure-dependent reflectance measurements were carried out on Na$_2$IrO$_3$ single crystals in the frequency range $200-20000\,$cm$^{-1}$ ($0.025-2.5\,$eV).
To obtain the reflectance ratio $R_\text{sd}$ at the sample-diamond interface, the intensity $I_\text{s,refl}(\omega)$ of the radiation reflected from the sample-diamond interface was measured.
As reference, the intensity $I_\text{ref,refl}(\omega)$ of the radiation reflected from the inner diamond air interface of the empty diamond anvil cell (DAC) was used.
The reflectance ratios were calculated according to $R_\text{sd}(\omega)=R_\text{dia} \cdot I_\text{s,refl}(\omega)/I_\text{ref,refl}(\omega)$ with $R_\text{dia}=0.167$, which was assumed to be independent of pressure \cite{Eremets.1992, Ruoff.1994}.
The reflectance ratios $R_\text{sd}$ were calibrated against the simulated $R_\text{sd}$ obtained from fitting the absolute reflectance outside the cell with the Lorentz model.
The calibrated spectra were then fitted with a large number of Lorentz terms (variational dielectric function \cite{Kuzmenko.2005}), using the RefFit software, to obtain the real part of the optical conductivity at various pressures.

A commercial Diacell{\textregistered} CryoDAC-Mega (almax-easylab) DAC was used for generating pressures up to 14$\,$GPa in the FIR range, while in the MIR, NIR and VIS range a custom-made Syassen-Holzapfel type \cite{Huber.1977} DAC was used for generating pressures up to $24\,$GPa.
Pressure was determined \textit{in situ} by using the Ruby luminescence technique \cite{Mao.1986}.
CsI powder served as quasi-hydrostatic pressure transmitting medium.

Since the samples are highly air sensitive \cite{Krizan.2014}, they were kept inside an Ar-filled glove box.
They were measured quickly after exposing them to air and were stored inside a vacuum desiccator in between measurements.
The pressure-dependent spectra were observed from a freshly cleaved sample, which was quickly loaded into the DAC.
The typical sample size used for pressure-dependent measurements amounted to $\approx 120\,\upmu$m$\times 120\,\upmu$m for the FIR range, and $\approx 90\,\upmu$m$\times 90\,\upmu$m for the MIR, NIR, and VIS ranges.

The pressure dependence of the lattice parameters was determined by single-crystal x-ray diffraction (XRD) measurements using synchrotron radiation at beamline ID15B at the ESRF (Grenoble, France).
The wavelength of the radiation was $0.4114\,${\AA}, and more than $300$ reflections were used to determine the crystal structure.
Diffraction data were analyzed using the CrysAlisPro software (Rigaku Oxford Diffraction, CrysAlisPro Software system. (1995 - 2015)), following the established protocols for the beamline \cite{Merlini.2013}.

Phonon frequencies at the $\Gamma$-point were obtained from density-functional (DFT) band-structure calculations using the internal procedure of \texttt{VASP}~\cite{vasp1,vasp2} that adopts finite displacements method for the calculation of phonons. The Perdew-Burke-Ernzerhof exchange-correlation potential for solids~\cite{Perdew.2008} was used along with the $k$-mesh of up to 64 points in the first Brillouin zone. Given variable magnetic ground states of the honeycomb iridates considered in this work, all calculations were performed for the ferromagnetic spin configuration. We have verified that collinear antiferromagnetic order changes phonon frequencies of Na$_2$IrO$_3$ by less than 1\,\%, and magnetic order has negligible effect on lattice dynamics. All calculations were performed on the DFT+$U$+SO level, because both correlations and spin-orbit coupling are integral to stabilizing the insulating state of honeycomb iridates~\cite{Foyevtsova.2013,Kim.2016}. The on-site Hubbard repulsion and Hund's exchange parameters of DFT+$U$+SO were fixed at $U_d=2$\,eV and $J_d=0.4$\,eV~\cite{Winter.2016}. Their variation systematically changes phonon frequencies by a few percent without affecting any of the conclusions presented below.

\section{Results and Discussion}

\subsection{Isoelectronic doping}

\begin{figure}
\centering
\includegraphics[width=0.45\textwidth]{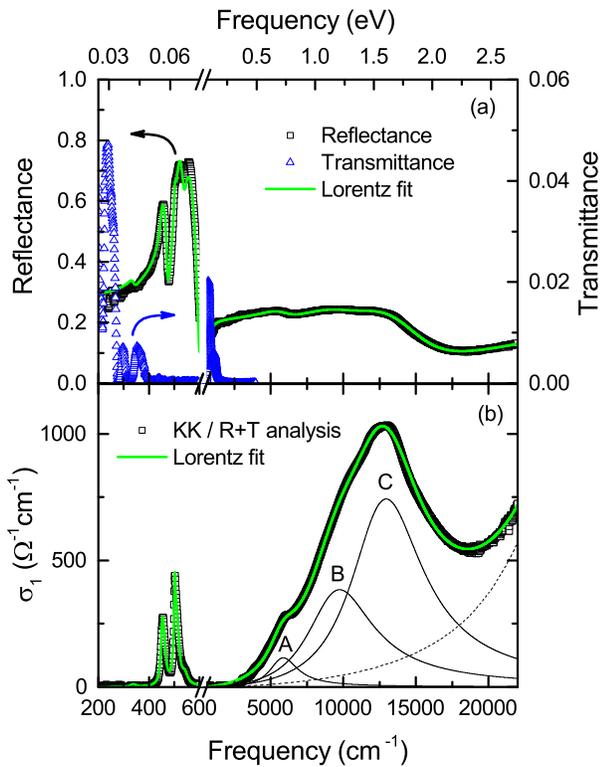}
\caption{(Color online)
(a) Reflectance (black squares) and transmittance (blue up triangles) spectra of Na$_2$IrO$_3$ under ambient conditions. The green line is the fit of the reflectance spectrum with Lorentz oscillators.
(b) Optical conductivity $\sigma_1$ using a combined Kramers-Kronig and R+T analysis, together with the fitting with Lorentz oscillators. The pronounced absorption feature at around $1.5\,$eV consists of three main contributions labeled A, B, and C. The dashed line indicates the charge-transfer excitations between the Ir $5d$ and O $2p$ orbitals.}\label{img.fig1}
\end{figure}

The reflectance and transmittance spectra of the measured Na$_2$IrO$_3$ crystal is depicted in Fig.~\ref{img.fig1} (a).
Consistent with the insulating character of the material \cite{Jin.2009, Mazin.2012, Mazin.2013, Kim.2016b, Comin.2012, Sohn.2013} the overall reflectivity is low, except for the frequency range $400-600\,$cm$^{-1}$ due to strong phonon excitations.
The sample is opaque in almost the whole studied frequency range, as given by the transmittance spectrum [see Fig.~\ref{img.fig1} (a)].
For frequencies between $600-2000\,$cm$^{-1}$ ($0.074-0.25\,$eV) and below $400\,$cm$^{-1}$ ($50\,$meV) we obtain a transmittance up to $5\,\%$.
The corresponding optical conductivity $\sigma_1$ [Fig.~\ref{img.fig1} (b)] shows a pronounced absorption feature at around $1.5\,$eV with an absorption onset at $\sim$340~meV, confirming earlier experimental reports \cite{Comin.2012, Sohn.2013}.
Above $\sim$2.5~eV charge-transfer excitations between the Ir $5d$ and O $2p$ orbitals contribute to the optical conductivity spectrum.

Fitting the pronounced absorption feature with the Lorentz model revealed three main contributions: The strongest Lorentz contribution labelled \textbf{C} is located at 1.6~eV and two weaker contributions \textbf{A} and \textbf{B} at $0.7$ and $1.2\,$eV, respectively.
These contributions can be ascribed to Ir $d-d$ transitions \cite{Sohn.2013, Kim.2016, Gretarsson.2013}.
Compared to the findings reported by Sohn \textit{et al.} \cite{Sohn.2013}, the contributions are slightly shifted in energy and the two additional Lorentzian peaks at 0.5 and 2.0~eV are not needed to obtain a good fit of our data.
Based on recent theoretical calculations, the three contributions \textbf{A}, \textbf{B}, and \textbf{C} can be ascribed to excitations between the relativistic $j_\text{eff}$=3/2 ($j_{3/2}$) and $j_\text{eff}$=1/2 ($j_{1/2}$) orbitals \cite{Li.2015,Kim.2016,Li.2017}.
Peak \textbf{C} can be ascribed to intersite $j_{3/2} \rightarrow j_{1/2}$ excitations, peak \textbf{B} to weaker intersite $j_{1/2} \rightarrow j_{1/2}$ excitations, and peak \textbf{A} to on-site $j_{3/2} \rightarrow j_{1/2}$ excitations.
These on-site excitations are expected to be weaker according to selection rules \cite{Li.2017}.

\begin{figure}
\centering
\includegraphics[width=0.45\textwidth]{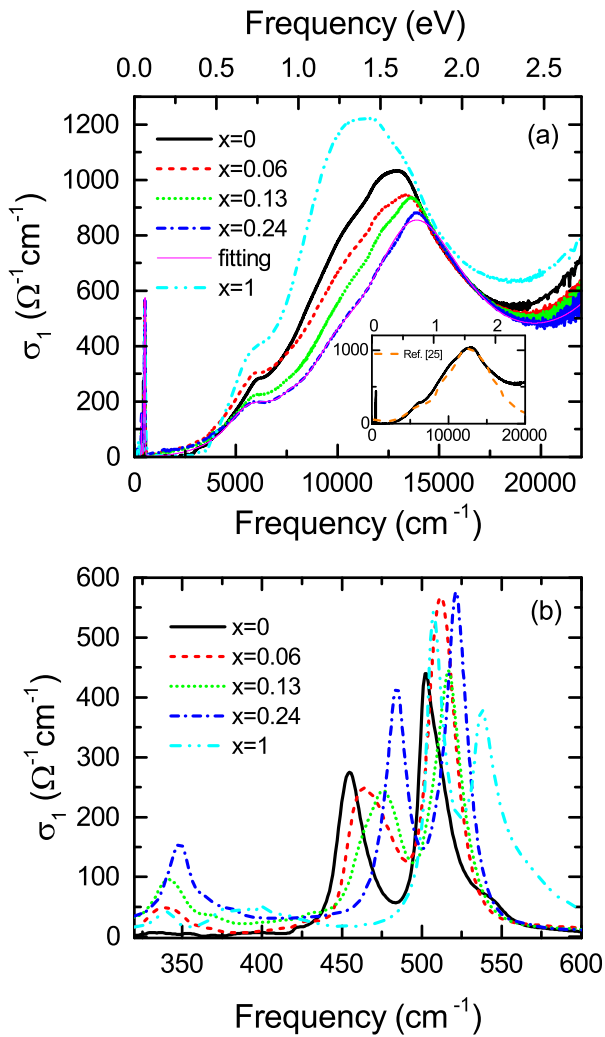}
\caption{(Color online) Optical conductivity $\sigma_1$ of (Na$_{1-x}$Li$_x$)$_2$IrO$_3$ for various Li contents $x$ (a) for the complete measured frequency range, and (b) for the phonon mode range. Inset of (a): Comparison between calculations of Ref. \cite{Kim.2016} and our result with parameters given in the text.}\label{img.fig2}
\end{figure}

The optical conductivity spectrum of $t^5_{2g}$ systems with a honeycomb lattice structure was theoretically investigated recently \cite{Kim.2016}, starting from a minimal microscopic model, which is able to capture the QMO band insulating limit as well as the relativistic Mott-insulating regime, taking into account the Coulomb repulsion $U$, Hund's coupling $J_\text{H}$, and SOC $\lambda$.
Kim et al. furthermore calculated the hole density $\overline{n}_{a1g}$ of the $a_{1g}$ quasimolecular band at the $\Gamma$ point as an indication for the QMO character of the material, which should be exactly one for the pure QMO state \cite{Kim.2016}.
Comparing the shape of our optical conductivity to their calculated spectra with the Kubo formula and with $J_\text{H}=1.6t$ ($t$ being the hopping parameter between adjacent Ir atoms), we find the best match for $\lambda$ at $1.6t$ and $U-3J_\text{H}=2.0t$.
Setting $t$ to a reasonable value of $t=0.31$~eV, which is slightly higher than $t\approx 0.27$~eV from various theoretical calculations \cite{Mazin.2012, Foyevtsova.2013, Yamaji.2014}, we find the three peak structure matching very well to our obtained spectrum as shown in the inset of Fig.~\ref{img.fig2} (a).
In comparison, setting $t\approx0.27$~eV as implied by Ref. \cite{Kim.2016} will make the spectrum with $U-3J_\text{H}=3.2t$ fit to our main peak \textbf{C} reasonable well, but peak \textbf{A} is then too much underestimated.
The hole density is estimated to be around $\overline{n}_{a1g}\approx0.35$.
Thus, our results indicate that Na$_2$IrO$_3$ is closer to a band insulating QMO state than the prediction from Kim et al., but still within the Mott-insulating regime \cite{Kim.2016}.

The effect of Li-doping on the optical conductivity of Na$_2$IrO$_3$ in the frequency range of the Ir $d$-$d$ transitions and the phonon modes is illustrated in Figs.\ \ref{img.fig2}(a) and (b), respectively.
Applying a simple Lorentz model with three contributions similar like for Na$_2$IrO$_3$ we find that the main peak {\bf C} shifts to higher energies with increasing Li content $x$ up to $x=0.24$ ($\sim$0.1~eV for $x=0.24$), which is the highest doping level achieved in our samples, whereas the lowest-energy contribution {\bf A} is stable regarding Li substitution, with only a slight tendency towards lower energy.
The overall spectral weight of the $d$-$d$ excitation band decreases with increasing doping level.
One furthermore notices a small spectral weight contribution below $\sim$340~meV for the Li-doped samples as compared to pure Na$_2$IrO$_3$.
According to calculations of Kim \textit{et al.}~\cite{Kim.2016}, the decrease in the spectral weight along with a shift of the main excitation \textbf{C} to higher energies can refer to an increase of the $(U-3J_\text{H})/t$ value, and thereby going more into the Mott-insulating regime,
which is quite surprising and will be discussed in section C.

For pure $\alpha$-Li$_2$IrO$_3$ we find a charge gap very similar in size ($\sim$340~meV) like for Na$_2$IrO$_3$, consistent with theoretical predictions \cite{Li.2015,Li.2017}. However, the optical conductivity spectrum does not follow the above-described trends for the Ir $d$-$d$ transitions, which we observe for the Li-doped Na$_2$IrO$_3$ compounds.
Compared to Na$_2$IrO$_3$ the overall spectral weight of the $d$-$d$ excitations is increased, with the main contribution centered at around 1.4~eV, i.e., at considerably lower energy as in Na$_2$IrO$_3$.
An increase in spectral weight between 0.7 and 1.5~eV was theoretically predicted for $\alpha$-Li$_2$IrO$_3$~\cite{Li.2015,Li.2017}.
In particular, in contrast to Na$_2$IrO$_3$ the direct metal-metal hopping was shown to be significant in the case of $\alpha$-Li$_2$IrO$_3$, which leads to additional spectral weight of $j_{1/2} \rightarrow j_{1/2}$ excitations located at around 1.1~eV.
This is in accordance with the calculations of Kim \textit{et al.}~\cite{Kim.2016} for Na$_2$IrO$_3$.
The increase of the spectral weight along with a shift of the main peak to lower energies is a signal for a reduce in the $(U-3J_\text{H})/t$ value, simply explained by an increase of the hopping parameter $t$ due to the decrease of lattice parameters.
This brings $\alpha$-Li$_2$IrO$_3$ closer to the QMO limit than Na$_2$IrO$_3$.

Furthermore, the honeycomb lattices of $A_2$IrO$_3$ are discussed in terms of Kitaev interactions \cite{Singh.2012, Knolle.2014, Chaloupka.2015, Chaloupka.2010, Winter.2016, Yamaji.2016}.
According to Li \textit{et al.}~\cite{Li.2017}, the Kitaev limit will be most closely approached by the (Na$_{1-x}$Li$_x$)$_2$IrO$_3$ compound with the lowest spectral weight near $\omega \approx 1.1\,$eV, where the spectral weight is contributed by $j_{1/2} \rightarrow j_{1/2}$ excitations.
They furthermore claim that the values of the corresponding direct metal-metal hopping integrals are directly related to the magnetic interactions and that the Kitaev limit will only be obtained, when the direct metal-metal hopping is small in comparison to the oxygen assisted intersite hopping.
Following this interpretation would identify the $24\,\%$ doped (Na$_{0.76}$Li$_{0.24}$)$_2$IrO$_3$ as the closest of the measured materials to the Kitaev limit signaled by the decrease of contribution \textbf{B}, while the enhanced direct metal-metal hopping signals that $\alpha$-Li$_2$IrO$_3$ should be less close to the Kitaev-limit as compared to Na$_2$IrO$_3$ \cite{Li.2017, Winter.2016}.
The proposed shift toward the Kitaev-limit upon Li substitution is consistent with the reduction in the $T_N$ from 15\,K at $x=0$ to 6\,K at $x=0.24$. On the other hand, at $x=1$ the $T_N$ of about 15\,K is recovered \cite{Manni.2014}.

\begin{figure}
\centering
\includegraphics[width=0.45\textwidth]{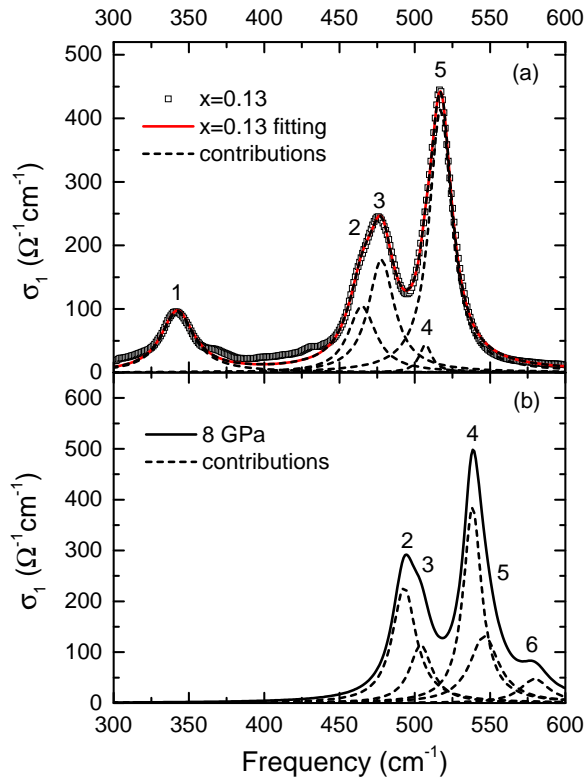}
\caption{(Color online) Optical conductivity $\sigma_1$ of (Na$_{1-x}$Li$_x$)$_2$IrO$_3$ in the phonon mode range: (a) For $x$=0.13 at ambient pressure together with the fitting with the Lorentz model and the Lorentzian contributions 1 to 5. (b) For $x$=0 at the pressure 8~GPa together with the Lorentzian contributions 2 to 6.}\label{img.fig3}
\end{figure}

Next we focus on the analysis of the phonon mode spectrum of Na$_2$IrO$_3$ and its changes with Li doping.
Experimentally we observe up to 6 phonon modes which are listed in Table~\ref{tab:phon_exp} and whose frequencies are plotted in Fig.~\ref{img.fig6} (c) as a function of Li doping level $x$.
With increasing Li content $x$ all observed phonon modes harden [see also Fig.~\ref{img.fig2}(b)].
The fit of the phonon mode spectrum with Lorentz oscillators is depicted in Fig. \ref{img.fig3}(a) for the Li concentration $x$=0.13 as an example, together with the labeling of the five phonon modes.
Phonon mode 1 is observed only in samples containing Li (Please note that for the pure Li compound, the phonon mode 1 is present but the absolute value of the optical conductivity could not be determined due to technical reasons, namely very low intensity of the transmission for the R+T analysis).
An additional mode 6 is observed only in the pure compounds Na$_2$IrO$_3$ and $\alpha$-Li$_2$IrO$_3$, but not in the doped compounds.

\begin{table}
\caption{\label{tab:phon_exp} Experimentally observed phonon frequencies in (cm$^{-1}$) for Na$_2$IrO$_3$, Na$_{1.52}$Li$_{0.48}$IrO$_3$, and $\alpha$-Li$_2$IrO$_3$ at ambient pressure.}
\begin{ruledtabular}
\begin{tabular}{cccc}

mode  & Na$_2$IrO$_3$ & (Na$_{0.76}$Li$_{0.24}$)$_2$IrO$_3$ & $\alpha$-Li$_2$IrO$_3$ \\
1 & -- & 347 & 387 \\
2 & 452 & 475 & 506 \\
3 & 460 & 484 & 512 \\
4 & 502 & 510 & 537 \\
5 & 511 & 522 & 540 \\
6 & 542 & -- & 566 \\

\end{tabular}
\end{ruledtabular}
\end{table}

DFT+$U$+SO calculations provide microscopic insight into $\Gamma$-point phonons probed in the optical measurements. The $C2/m$ symmetry of the crystal structure allows 18 infrared-active modes that are split into 7 modes of the $A_u$ symmetry and 11 modes of the $B_u$ symmetry. Their frequencies and natures are listed in Table~\ref{tab:phonons}, where we restrict ourselves to modes above 250\,cm$^{-1}$, because low-energy modes are less characteristic and could not be probed in our experiment.

\begingroup
\begin{table}
\caption{\label{tab:phonons}
Calculated $\Gamma$-point phonon frequencies (in\,cm$^{-1}$) for Na$_2$IrO$_3$ ($x=0$) at ambient pressure, compressed Na$_2$IrO$_3$ (comp., with lattice parameters of the $x=0.24$ sample but no Li substitution),
Na$_{1.5}$Li$_{0.5}$IrO$_3$ ($x=0.25$, with lattice parameters of the $x$=0.24 sample and placing Li atoms into the center of the Ir hexagons) at ambient pressure, and $\alpha$-Li$_2$IrO$_3$ ($x=1.0$) at ambient pressure. For each mode, the atoms with largest displacements are listed.
}
\begin{ruledtabular}
\begin{tabular}{ccccc|ccc}

mode  &    & Na$_2$IrO$_3$ & & Na$_2$IrO$_3$ & $x$=$0.25$ & & $\alpha$-Li$_2$IrO$_3$ \\
 & & & & comp. & & & \\
c1 & $A_u$ &  544 & O & 567 & 570 & O & 579 \\
c2 & $A_u$ &  503 & Ir--O & 525 & 532 & Ir--O--Li & 528 \\
c3 & $B_u$ &  502 & Ir--O & 518 & 535 & Ir--O--Li & 542 \\
c4 & $B_u$ &  497 & Ir--O & 515 & 525 & Ir--O--Li & 527 \\
c5 & $A_u$ &  454 & Ir--O & 506 & 485 & Ir--O--Li & 501 \\
c6 & $B_u$ &  453 & Ir--O & 507 & 484 & Ir--O--Li & 503 \\
c7 & $B_u$ &  425 & Na--O & 469 & 369 & Li & 406 \\
c8 & $A_u$ &  413 & Na--O & 463 & 361 & Li & 385 \\
c9 & $B_u$ &  379 & Na--O &453 & 346 & Li & 392 \\
c10 & $B_u$ & 288 & Na & 299 & 344 & Li & 331 \\
c11 & $A_u$ & 269 & Na & 289 & 336 & Li & 337 \\
c12 & $B_u$ & 267 & Na & 283 & 310 & Li & 314 \\
\end{tabular}
\end{ruledtabular}
\end{table}
\endgroup

In Na$_2$IrO$_3$, only the high-frequency mode c1 at 544\,cm$^{-1}$ (mode 6 in Fig.~\ref{img.fig6}c) is a purely oxygen-based vibration. Five modes c2--c6 found between 453 and 503\,cm$^{-1}$ entail significant contributions of Ir and can be seen as collective vibrations of the Ir--O framework. Experimentally, we resolve only 4 modes in this frequency range (modes 2--5 in Fig.~\ref{img.fig6}c), because some of the modes are nearly degenerate. Three further modes c7--c9 found between 379 and 425\,cm$^{-1}$ and not observed experimentally, presumably due to their low oscillator strengths, are Na--O vibrations, whereas further infrared-active modes positioned below 288\,cm$^{-1}$ are dominated by Na atoms.

We explored the effect of Li doping by using lattice parameters of the $x=0.24$ sample and placing Li atoms into the center of the Ir hexagons (see phonon modes in Table~\ref{tab:phonons}, column $x$=0.25). The high-energy mode c1 preserves its oxygen-based nature. However, the modes c2, c5, and c6 that previously did not involve Na atoms, feature now comparable contributions of Ir and Li. The spectrum at lower energies changes entirely. No equivalent of the modes c7--c9 can be found, and instead Li-based vibrations dominate the spectrum below 369\,cm$^{-1}$. This elucidates the origin of mode 1 that was observed in the Li-doped samples but not in the pure Na$_2$IrO$_3$. Moreover, the phonon spectrum at $x=0.25$ bears strong similarities to the one of $\alpha$-Li$_2$IrO$_3$ in terms of the nature of phonon modes, because both Li and Ir atoms contribute to the modes c2--c6. No infrared-active modes are seen between 406 and 501\,cm$^{-1}$, whereas below 406\,cm$^{-1}$ multiple Li-based vibrations occur.

\begin{figure}
\centering
\includegraphics[width=0.48\textwidth]{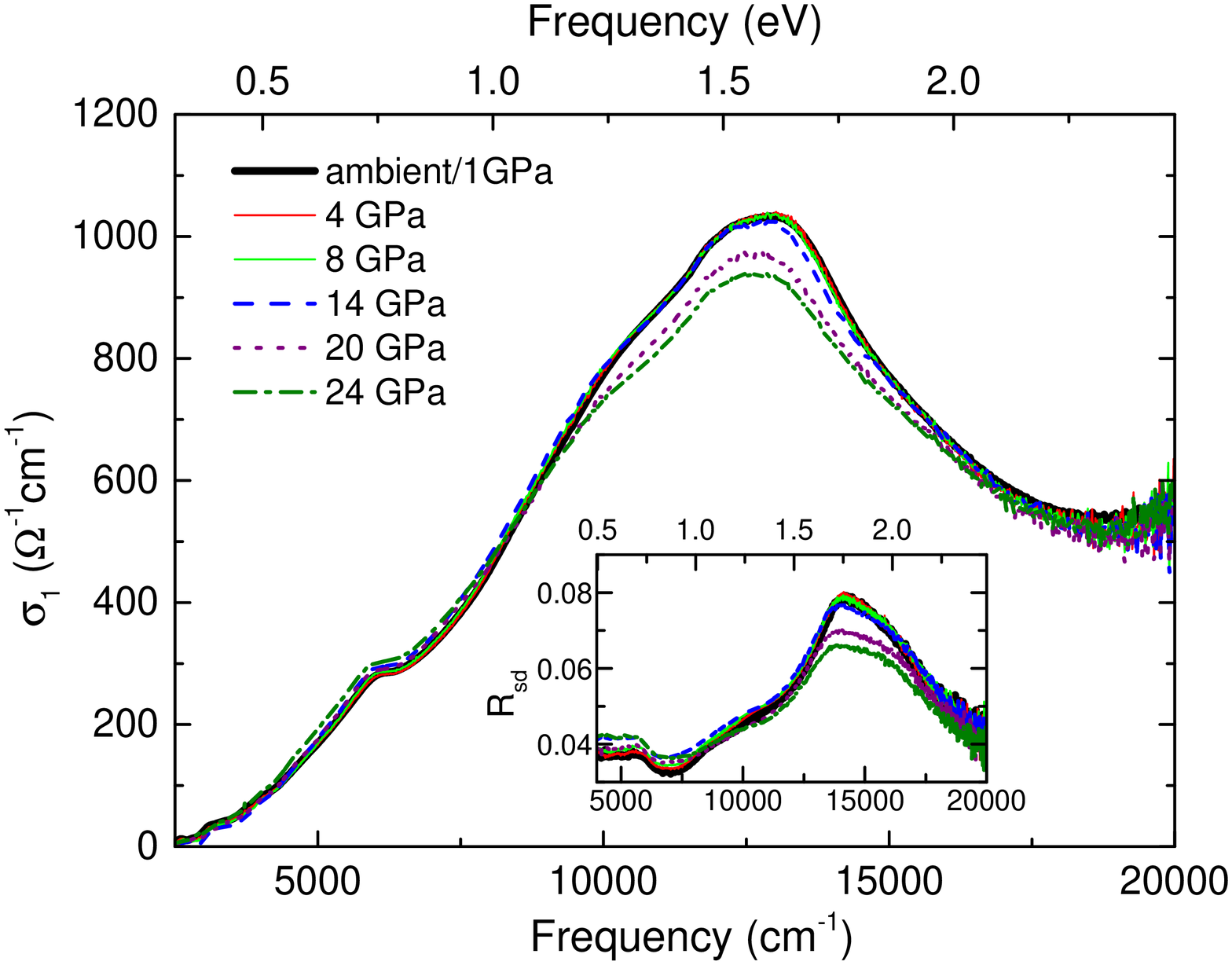}
\caption{(Color online) Pressure-dependent conductivity spectra in the frequency range of the $d$-$d$ excitations of Na$_2$IrO$_3$ up to $24\,$GPa. Inset: Pressure-dependent reflectivity spectra.}\label{img.fig4}
\end{figure}

Replacing atoms in a solid by isoelectronic atoms with smaller covalent radius is expected to lead to a chemical pressure effect without any charge doping.
Applying this general rule to Na$_2$IrO$_3$, one would expect a significant reduction of the lattice parameters in Na$_2$IrO$_3$ by substituting Na with smaller Li atoms.
It has been shown that the changes in the lattice parameters in Na$_2$IrO$_3$ induced by Li doping are specific \cite{Manni.2014}: While the lattice parameters $a$ and $b$ are strongly decreased at the same rate by increasing Li content $x$ (leading to an almost undistorted honeycomb Ir structure up to $x\leq 0.25$), the lattice parameter $c$ remains almost unchanged.
Hence, the $c$/$a$ lattice parameter ratio increases with increasing doping $x$.
This experimental finding was explained by the fact that Li atoms preferentially occupy the Na sites within the Ir honeycomb layers for $x\leq 0.25$, consistent with theoretical calculations \cite{Manni.2014}.
The Na layers between the honeycomb layers remain basically unaffected for such low doping concentrations, and hence no effective pressure is applied along the $c$ axis.
The effect of external, hydrostatic pressure on the physical properties of Na$_2$IrO$_3$ is expected to be quite different, and hence a direct comparison with the effect of Li doping on the physical properties like crystal structure and optical response is appealing.
Therefore, we studied the optical properties and crystal structure of Na$_2$IrO$_3$ single crystals as a function of external pressure.

\subsection{External pressure}

\begin{figure}
\centering
\includegraphics[width=0.4\textwidth]{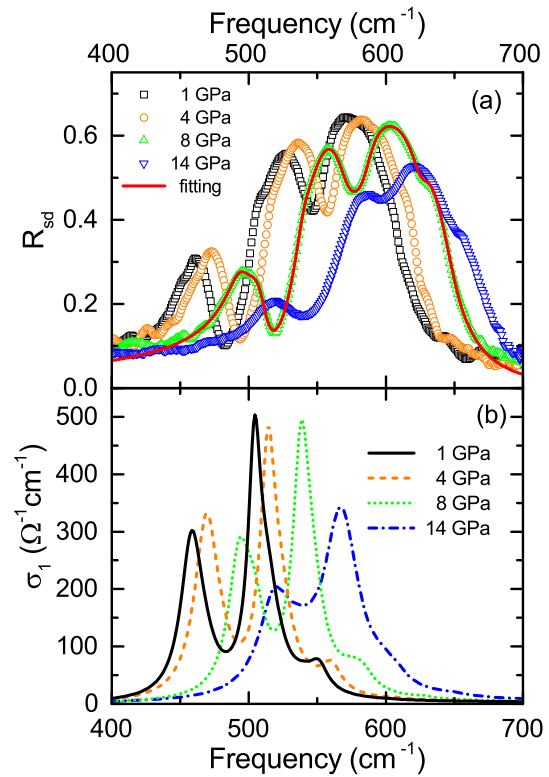}
\caption{(Color online) (a) Pressure-dependent reflectivity spectra $R_{sd}$ of Na$_2$IrO$_3$ in the phonon mode range for selected pressures. As an example, the fitting of the reflectivity spectrum at $8\,$GPa with the Lorentz model is shown. The corresponding real part of the optical conductivity is plotted in (b). With increasing pressure all phonon modes harden and show a damping above $8\,$GPa.}\label{img.fig5}
\end{figure}

The pressure-dependent optical conductivity spectra of Na$_2$IrO$_3$ are shown in Fig. \ref{img.fig4} in the range of the $d$-$d$ excitations and in Fig. \ref{img.fig5} (b) for the phonon energy range for some selected pressures.
The reflectance ratios $R_{sd}$ are included in the corresponding figures.
In the range of the $d$-$d$ excitations we do not observe a significant change of the optical conductivity up to $8\,$GPa.
For pressures above $8\,$GPa we observe a significant loss of spectral weight for the main excitation along with a slight shift of the three main peaks to lower energy.
An additional contribution appears at around $2.0\,$eV, consistent with earlier studies \cite{Sohn.2013}, which is more clearly observed for the higher pressures.
Overall, the optical excitations in Na$_2$IrO$_3$ are quite robust regarding external pressure.
These findings suggest that the $(U-3J_\text{H})/t$ value is not significantly affected for pressures up to $8\,$GPa and is only slightly decreased for higher pressures.

In Fig. \ref{img.fig5} (b) the reflectivity spectra of Na$_2$IrO$_3$ are depicted in the phonon mode range for selected pressures. In order to obtain the phonon frequencies as a function of pressure, we performed a fitting with Lorentz functions, taking into account the sample-diamond interface and the fitting of the ambient-pressure data as extrapolation \footnote{From the Fresnel equations one obtains: $R_{s-d}=\left| (\sqrt{\epsilon_s}-\sqrt{\epsilon_d})/(\sqrt{\epsilon_s}+\sqrt{\epsilon_d})\right|^2$ with $\epsilon_d$ being the dielectric function of diamond and $\epsilon_s$ the dielectric function of the measured sample.}.
As an example, the fitting of the reflectance spectrum at $8\,$GPa is shown in Fig.~\ref{img.fig5}(a).
The so-obtained optical conductivity spectra in the phonon mode range are depicted in Fig. \ref{img.fig5} (b).
We find five major phonon modes in the range $400-600\,$cm$^{-1}$, labeled with $2-6$ according to Fig. \ref{img.fig3}(b), consistent with the ambient-pressure data.
Similar to the free-standing measurements phonon mode 1 has an extremely small oscillator strength and is accessible only by R+T analysis.
Since the oscillator strength of this mode is not increased with pressure, we will not discuss it further.
Like in the ambient-pressure data, we obtain two double-peak structures, with a small contribution of mode 6.
There is an additional very weak contribution in the high-energy range of the phonon mode region, which is not considered in the following due to its extremely small oscillator strength.

\begin{figure}
\centering
\includegraphics[width=0.46\textwidth]{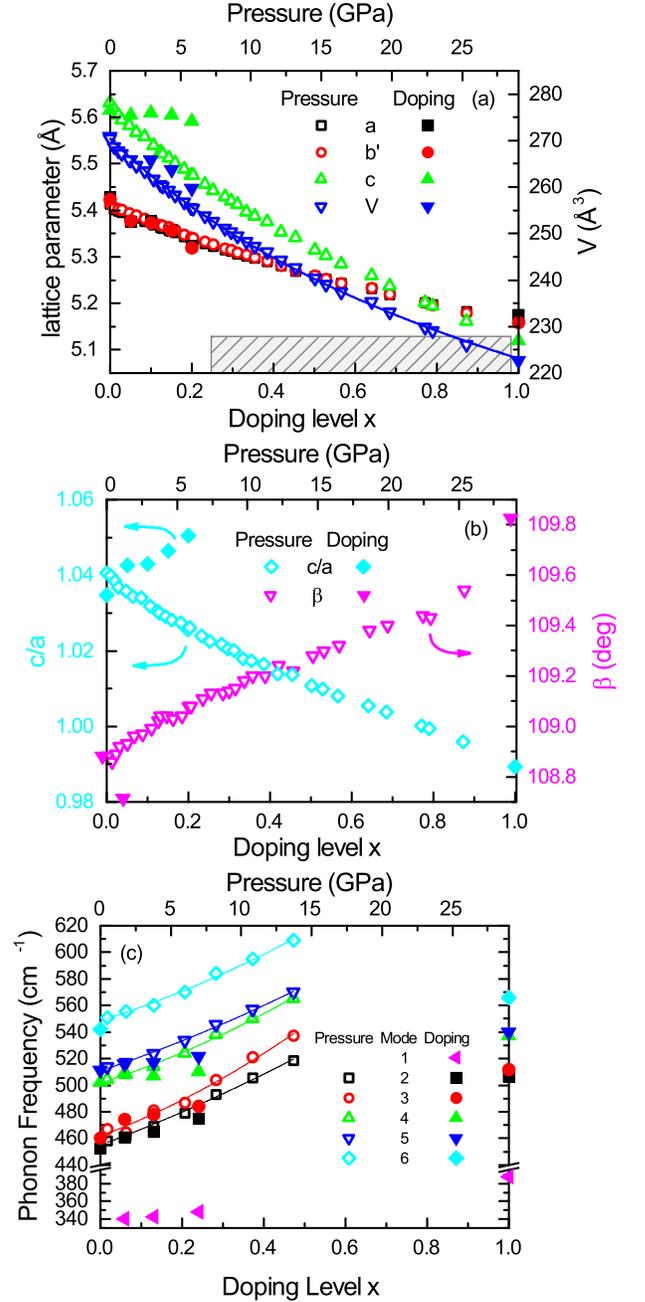}
\caption{(Color online) (a) and (b): Lattice parameters $a$, $b':=b/\sqrt{3}$, $c$, $\beta$, unit cell volume V, and $c/a$ ratio of Na$_2$IrO$_3$ as a function of external pressure (open symbols) and Li doping level $x$ (filled symbols). The results for doping levels $x$=$0.05-0.2$ and $x=1$ are extracted from Refs.~\onlinecite{Manni.2014} and \onlinecite{Freund.2016}, resp.. The pressure scale is adjusted to the doping level $x$ according to the evolution of the lattice parameters $a$ and $b$. The volume V is fitted by the Murnaghan equation of states as discussed in the text. The hatched area in (a) marks the miscibility gap for doping levels 0.25$<$$x$$<$1 \cite{Manni.2014}.
(c) Phonon modes frequencies as a function of external pressure and Li doping level $x$, keeping the pressure scaling as adjusted for the lattice parameters in (a) and (b). The solid lines are parabolic fits to the pressure-dependent mode frequencies.}\label{img.fig6}
\end{figure}

The pressure-dependent phonon frequencies are plotted in Fig. \ref{img.fig6} (c).
We observe a monotonic hardening of all five phonon modes with increasing pressure, in agreement with the DFT results (Table II).
Additionally, a broadening of the modes occurs for pressures above $8\,$GPa (see Fig.~\ref{img.fig5}).
Generally, the damping of phonon modes indicates an increasing metallic character of a material, which is, however, not revealed by the pressure dependence of the overall reflectivity spectrum.
In other words, it is a signal for an only small decrease of the $(U-3J_\text{H})/t$ value, consistent with the results from the $d$-$d$ excitations.

Next we discuss the effect of external pressure on the crystal structure of Na$_2$IrO$_3$ based on pressure-dependent XRD measurements carried out on single crystals.
In Figs. \ref{img.fig6}(a) and (b) we show the evolution of the lattice parameters $a$, $b':=b/\sqrt{3}$, $c$, $\beta$, unit cell volume V, and $c/a$ ratio of Na$_2$IrO$_3$ as a function of external pressure in comparison to the evolution with Li doping level $x$, as extracted from Ref.~\onlinecite{Manni.2014}.
The parameter $b'$ allows a more direct comparison between the lattice parameters $a$ and $b$ regarding the pressure evolution.
The pressure scale is adjusted to the doping level $x$ for $x\leq 0.24$ samples using $a$ and $b'$, and extended to $x=1$ that would correspond to the external pressure of 29~GPa.

With increasing pressure, the lattice parameters $a$, $b'$, and $c$ decrease monotonically, with the strongest effect observed for the lattice parameter $c$ perpendicular to the Ir honeycomb layers.
The effect of pressure on the lattice parameter $c$ is approximately doubled compared to the $a$ and $b'$ parameters.
Hence, the highest compressibility of Na$_2$IrO$_3$ is found along the $c$-axis, which is illustrated by the $c/a$ ratio that decreases from $1.04$ to $0.99$ [see Fig. \ref{img.fig6}(b)]. This can be ascribed to the two-dimensional nature of the Ir--O framework. The two in-plane lattice parameters $a$ and $b'$ are affected in a very similar manner, which excludes any pressure-induced distortion of the Ir hexagons.
The angle $\beta$ increases monotonically with increasing pressure [see Fig. \ref{img.fig6}(b)].
Furthermore, there is no sign of a pressure-induced structural phase transition within the measured pressure range.

In Fig. \ref{img.fig6}(a), the unit cell volume $V$ is plotted as a function of pressure $p$, as calculated from the pressure-dependent lattice parameters, together with the fit according to the Murnaghan equation \cite{Murnaghan.1944},
\begin{equation}
V(p)=V_0\cdot[(B_0'/B_0)\cdot p +1]^{-1/{B_0'}} \quad ,
\end{equation}
with the bulk modulus $B_0=-\mathrm{d} p/ \mathrm{d} \ln V$ and its first derivative $B_0'$ at ambient pressure.
From the fitting we obtained $B_0=89.9 \pm 1.2\,$GPa and $B_0'=5.0 \pm 0.2$.
In comparison, for pyrochlore iridates like Eu$_2$Ir$_2$O$_7$ and similar compounds like Eu$_2$Sn$_2$O$_7$ bulk moduli $B_0$ in the range $166$ to $285\,$GPa with $B_0'$ in the range $3.8$ to $28$ are reported.\cite{Clancy.2016, Zhao.2016, Zhang.2010, Saha.2009}
For the perovskite iridate Sr$_2$IrO$_4$ $B_0=174\pm 5$\,GPa and $B_0'=4.0 \pm 0.7$ is found \cite{Haskel.2012}, while for the stripy-honeycomb $\gamma$-Li$_2$IrO$_3$ $B_0$ around $130\,$GPa was reported recently \cite{Breznay.2017}.
Therefore, the bulk modulus for the honeycomb iridate Na$_2$IrO$_3$ is smaller than for pyrochlore, perovskite, and stripy-honeycomb iridates, i.e., Na$_2$IrO$_3$ is more compressible.
This can be attributed to the two-dimensional, layered character of its crystal structure, which can be relatively easily compressed along the $c$ axis, as illustrated by the pressure dependence of the lattice parameter $c$ [see Figs. \ref{img.fig6}(a)+(b)].
This anisotropic compressibility is well in line with the hierarchy of phonon modes.
The Na--O bonds responsible for the compression along $c$ are softer than their Ir--O counterparts, which determine the compressibility in the $ab$ plane (Table~\ref{tab:phonons}).
Comparing the bulk modulus of Na$_2$IrO$_3$ with typical layered compounds like Bi$_2$Te$_3$ ($B_0=22-28\,$GPa and $B_0'=13.8-17.1$ for low pressures and $B_0=36-38\,$GPa and $B_0'=4.6-5.5$ for higher pressures)~\cite{Jenkins.1972, Nakayama.2009, Polian.2011} or graphite ($B_0=33.8-39\,$GPa and $B_0'=8.9-10$)~\cite{Blakslee.1970, Hanfland.1989, Zhao.1989, Reich.2002}, however, reveals that Na$_2$IrO$_3$ is much less compressible.

\subsection{Comparison between isoelectronic doping and external pressure}

The optical conductivity identifies qualitative differences between the effects of partial Li substitution ($x\leq 0.24$), full Li substitution (pure $\alpha$-Li$_2$IrO$_3$), and external pressure. With both increasing Li content (up to $x$=0.24) and increasing pressure above $8\,$GPa, the spectral weight in the range of the $d$-$d$ excitations decreases. In the case of doping, this decrease is stronger, and also a shift of the main excitation to higher energies is observed. On the contrary, full substitution with Li leads to an increase of the spectral weight along with the shift of the main excitations towards lower energies. We interpret these differences in terms of the Mott-vs-QMO picture and in terms of the proximity to the Kitaev-limit. The partial Li substitution turns out to be most promising for approaching the Kitaev limit and, concurrently, enhances the Mott-insulating character. Full Li substitution goes in exactly the opposite direction by taking the system further away from the Kitaev limit and toward the QMO state. The effect of external pressure is relatively mild. At pressures below 8-10\,GPa, which are feasible for low-temperature magnetism studies, no visible changes in the optical conductivity occur.

These dissimilar trends are not immediately obvious from the structural perspective. Indeed, the physics of the hexagonal iridates is largely confined to the Ir--O layers, and the compression in the $ab$ plane is about the same in all three cases (Fig.~\ref{img.fig6}~(a)). Moreover, $\alpha$-Li$_2$IrO$_3$ may even be viewed as Na$_2$IrO$_3$ compressed to approximately 29$\,$GPa, regarding all three lattice parameters. On the other hand, it seems crucial that partial Li substitution leaves the $c$ parameter nearly unchanged, in stark contrast to the effects of external pressure and full Li substitution. Given that oxygen-assisted Ir--O--Ir hopping plays a central role in the physics of Na$_2$IrO$_3$~\cite{Li.2017}, we believe that not only the compression in the $ab$ plane, but also a change in the relative positions of Ir and O is crucial here. Indeed, oxygen-mediated hoppings largely depend on the Ir--O--Ir angle, which, in turn, should be affected by the $c$ parameter.

The difference between the fully substituted $\alpha$-Li$_2$IrO$_3$ and Na$_2$IrO$_3$ compressed to 29\,GPa is more subtle and can't be explained by the evolution of lattice parameters, but it becomes clearer when we consider the evolution of phonon modes. In general, one expects that compression of the structure caused by external pressure or isoelectronic doping with smaller Li atoms increases the phonon frequencies, because atoms get closer to each other, and individual bonds harden. In the case of Li doping, the introduction of lighter Li atoms is a concurrent effect that should increase phonon frequencies too. Indeed, Fig.~\ref{img.fig6}~(c) demonstrates steady increase in the experimental phonon frequencies, whereas the computed frequencies for c1--c6 increase as well (compare different columns in Table~\ref{tab:phonons}).
This trend no longer holds for the modes c7--c9, which shift to higher frequencies upon compression of Na$_2$IrO$_3$ (see column ``Na$_2$IrO$_3$ comp.'' in Table~\ref{tab:phonons})
but to lower frequencies upon partial or full Li substitution in pure $\alpha$-Li$_2$IrO$_3$ (columns $x$=0.25 and $x$=1 in Table~\ref{tab:phonons}). It becomes even more counter-intuitive when the nature of these phonon modes is considered. With heavier Na atoms replaced by lighter Li, one would hardly expect the softening of c7--c9 that occurs in $\alpha$-Li$_2$IrO$_3$. This shows that, despite similar lattice dimensions, $\alpha$-Li$_2$IrO$_3$ can not be viewed as an ambient-pressure analog of Na$_2$IrO$_3$ compressed to 29\,GPa.

Altogether, we see that the three effects considered in this study (partial and full Li substitution and external pressure) are all dissimilar, and among them partial Li substitution looks most promising for enhancing the Kitaev interaction term.

\section{Conclusion}

In summary, from investigating (Na$_{1-x}$Li$_x$)$_2$IrO$_3$ single crystals for doping levels $x${$\leq$}0.24 and $x$=1 by optical spectroscopy at ambient conditions we find that all measured compounds are relativistic Mott insulators although being close to the quasimolecular orbital regime.
Isoelectronic doping of Na$_2$IrO$_3$ by Li up to the doping level of $x=0.24$ brings the material closer to the Kitaev limit and more into the Mott-insulating regime.
From an experimental point of view all observed phonon modes harden with increasing Li content $x$ due to the chemical pressure effect, while our DFT+SOC+U calculations show that their nature is changed even for low dopings.
Full substitution of Na by Li atoms does not follow the above trends: in $\alpha$-Li$_2$IrO$_3$, the intersite $j_{1/2} \rightarrow j_{1/2}$ excitations appear to be enhanced according to the optical conductivity spectrum.
Furthermore, our data suggest that $\alpha$-Li$_2$IrO$_3$ is less close to the Kitaev limit and closer to the quasimolecular orbital regime than the Li doped samples.

Pressure-dependent optical and x-ray diffraction measurements on Na$_2$IrO$_3$ single crystals enabled a comparison of the effect of Li doping with external pressure. With increasing pressure, the lattice parameters of Na$_2$IrO$_3$ decrease monotonically, with the largest effect found for the parameter $c$ perpendicular to the honeycomb layers.
The lattice parameters of $\alpha$-Li$_2$IrO$_3$ are consistent with the pressure-induced changes of the lattice parameters of Na$_2$IrO$_3$ at 29~GPa, whereas for doping levels $x\leq 0.24$ the $c$ lattice parameter remains unchanged.
With increasing pressure all phonon modes harden, but their nature remains unchanged in contrast to the effect of Li doping.
The effect of pressure on the Ir $d$-$d$ excitations is much less pronounced as compared to Li substitution, and we found out, that partial Li substitution is the most suitable in tuning the system towards the Kitaev limit.

\vspace*{1em}
\begin{acknowledgments}
We thank Roser Valent{\'\i}, Steve Winter, and Ying Li for the fruitful discussions and the ESRF, Grenoble, for the provision of beamtime. The LA-ICP-MS measurements were performed by Dr. Andreas Kl\"ugel, Department of Geoscience, University of Bremen. This work was financially supported by the Federal Ministry of Education and Research (BMBF), Germany, through Grant No. 05K13WA1 (Verbundprojekt 05K2013, Teilprojekt 1, PT-DESY). PG acknowledges financial support by the Deutsche Forschungsgemeinschaft (DFG) through TRR 80 and  SPP 1666. AJ acknowledges support from the DFG through Grant No. JE 748/1. AAT acknowledges financial support from the Federal Ministry for Education and Research via the Sofja-Kovalevskaya Award of Alexander von Humboldt Foundation.
\end{acknowledgments}

\end{document}